\documentclass{PoS}

\usepackage{graphicx}
\usepackage{amssymb}
\usepackage{amsmath}

\makeatletter
\def\simleq{\mathrel{\mathpalette\gl@align<}}
\def\simgeq{\mathrel{\mathpalette\gl@align>}}
\def\gl@align#1#2{\lower.6ex\vbox{\baselineskip\z@skip\lineskip\z@
     \ialign{$\m@th#1\hfill##\hfil$\crcr#2\crcr\sim\crcr}}}
\makeatother

\newcommand{\Luscher}{L\"uscher}

\title{Cutoff effects on lattice nuclear forces}

\ShortTitle{Cutoff effects on lattice nuclear forces}

\author{\speaker{Takumi Doi}\\%
Theoretical Research Division, Nishina Center, RIKEN, Wako 351-0198, Japan\\
E-mail: \email{doi@ribf.riken.jp}}

\author{for HAL QCD Collaboration}

\author{\includegraphics[width=.30\textwidth]{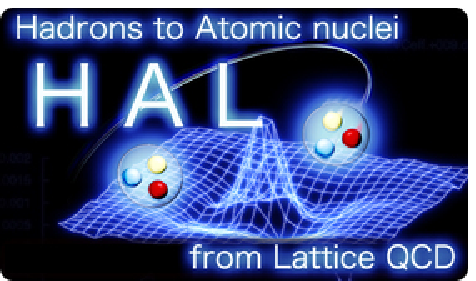}}

\abstract{
We present a lattice QCD study for 
the cutoff effects on nuclear forces.
Two-nucleon forces are determined
from Nambu-Bethe-Salpeter (NBS) wave functions
using the HAL QCD method.
Lattice QCD simulations are performed 
employing $N_f = 2$ clover fermion configurations
at three
lattice spacings of $a = 0.108, 0.156, 0.215$ fm
on a fixed physical volume of $L^3\times T \simeq (2.5 {\rm fm})^3 \times 5 {\rm fm}$
with a large quark mass corresponding to $m_\pi \simeq 1.1$ GeV.
We observe that while the discretization artifact appears 
at the short range part of potentials,
it is suppressed at the long distance region.
The cutoff dependence of the phase shifts and scattering length 
is also presented.
}

\FullConference{31st International Symposium on Lattice Field Theory - LATTICE 2013\\
		July 29 - August 3, 2013\\
		Mainz, Germany}

\begin{document}

\section{Introduction}
\label{sec:intro}

Nuclear forces,
the interactions among nucleons,
serve as the cornerstone in nuclear physics.
While they have been traditionally determined through the 
scattering experiments,
theoretical understanding of them from the fundamental theory, 
Quantum Chromodynamics (QCD), has not been established yet. 
In the last several years, substantial effort has been devoted
to determine nuclear interactions using lattice QCD simulations.
A conventional approach 
is to calculate an energy spectrum of 
a two-nucleon (2N) system on the lattice
and convert it to the scattering phase shift 
at the corresponding energy
through the \Luscher's finite volume formula~\cite{Luscher:1990ux}.
Latest lattice QCD results in this approach are given in, e.g., Refs.~\cite{Yamazaki:2012hi,Beane:2013br}.

Recently, a novel approach was proposed
to determine nuclear forces on the lattice~\cite{Ishii:2006ec,Aoki:2009ji}.
In this approach, now called as the HAL QCD method,
nuclear forces (or potentials)
are directly obtained from Nambu-Bethe-Salpeter (NBS) wave functions 
calculated in lattice simulations.
In particular, the extension to the ``time-dependent'' HAL QCD method
has an advantageous feature that 
energy-independent (non-local) potentials can be extracted
without replying on the ground state saturation~\cite{HALQCD:2012aa}.
Once potentials are obtained, scattering parameters such as phase shifts and scattering length
can be calculated by solving the Schr\"odinger equation in infinite volume
at arbitrary energies below the inelastic threshold.
Resultant (parity-even) nuclear potentials 
are found to have desirable features such as
attractive wells at long and medium
distances  and central repulsive cores at short distance.
The method has been successfully applied to more general hadron interactions,
such as hyperon interactions, 2N spin-orbit forces
and three-nucleon forces~\cite{Nemura:2008sp,Inoue:2010hs,Inoue:2010es,Murano:2011nz,Inoue:2011ai,Murano:2013xxa,Doi:2011gq}.
See Ref.~\cite{Aoki:2012tk} for a recent review.

Toward the quantitative determination of nuclear forces,
systematic uncertainties in lattice simulations should be carefully examined,
such as the effect of discretization artifact.
There have been, however, 
no work which performs the continuum extrapolation on nuclear interactions
in \Luscher's method nor in the HAL QCD method.
The aim of this work is to perform the first systematic study 
for the lattice cutoff dependence of nuclear interactions.
In particular, in the HAL QCD method, 
it could be easier to understand how the discretization artifact,
which has intrinsically short-range nature,
affects the lattice QCD results
since the spacial information remains in this method.
It is also interesting to examine 
how the characteristic feature in lattice nuclear potentials
such as repulsive cores at short distance
are stable against changing the lattice cutoff.
Therefore, in this study, we calculate nuclear potentials
in the HAL QCD method at three different lattice spacings,
with the other lattice 
parameters such as physical volume size and quark masses fixed.
Solving the Schr\"odinger equation with obtained potentials,
the cutoff dependence of scattering phase shifts and scattering 
length is also studied.


\section{Formalism}
\label{sec:formalism}

We briefly explain the framework of the HAL QCD method~\cite{Aoki:2009ji,HALQCD:2012aa,Aoki:2012tk}.
We consider the (equal-time) NBS wave function in the center-of-mass frame,
\begin{eqnarray}
\phi_W(\vec{r}) \equiv \langle 0 | N(\vec{r},0) N(\vec{0},0) | 2N, W \rangle_{\rm in} ,
\end{eqnarray}
where 
$N$ is the nucleon operator and
$|2N, W \rangle_{\rm in}$ denotes the asymptotic in-state of the 2N system 
at the total energy of $W = 2\sqrt{k^2+m_N^2}$
with the nucleon mass $m_N$ and the relative momentum $k \equiv |\vec{k}|$,
and we consider the elastic region,  $W < W_{\rm th} = 2m_N + m_\pi$.
For simplicity,
we omit other quantum numbers such as spinor/flavor indices.
%
%
%
The most important property of the NBS wave function is that
it has a desirable asymptotic behavior%
~\cite{Luscher:1990ux, Aoki:2009ji, Lin:2001ek, Aoki:2005uf, Ishizuka:2009bx},
%
\begin{eqnarray}
\phi_W (\vec{r}) \propto \frac{\sin(kr-l\pi/2 + \delta_l^W)}{kr}, 
\quad 
r \equiv |\vec{r}| \rightarrow \infty,
\end{eqnarray}
where 
$\delta_l^W$ is the scattering phase shift
with the orbital angular momentum $l$.
Exploiting this feature,
we define the (non-local) 2N potential, $U(\vec{r},\vec{r}')$,
through the following Schr\"odinger equation,
\begin{eqnarray}
%
H_0 \phi_W(\vec{r})
+ \int d\vec{r}' U(\vec{r},\vec{r}') \phi_W(\vec{r}')
= E_W \phi_W(\vec{r}) ,
\label{eq:Sch_2N:tindep}
\end{eqnarray}
where 
$H_0 = -\nabla^2/(2\mu)$ and
$E_W = k^2/(2\mu)$ with the reduced mass $\mu = m_N/2$.
It is evident that
$U(\vec{r},\vec{r}')$
defined in this way 
is faithful to the phase shift by construction.
%
In addition,
it has been proven that 
one can construct $U(\vec{r},\vec{r}')$
in an energy-independent way~\cite{Aoki:2009ji,Aoki:2012tk}.
These points guarantee that once potentials are obtained, 
one can determine the phase shifts at arbitrary energies below the inelastic threshold.
Furthermore, it is found that the energy-independence of potentials can be exploited
to make a reliable determination of potentials and phase shifts~\cite{HALQCD:2012aa}.
In fact, while the original equation (\ref{eq:Sch_2N:tindep}) requires the determination of 
$\phi_W$ for each energy $W$, by e.g., ground state saturation,
one can show that the same $U(\vec{r},\vec{r}')$ can be extracted
from the``time-dependent'' Schr\"odinger equation
\begin{eqnarray}
%
H_0 \psi(\vec{r},t)
+ \int d\vec{r}' U(\vec{r},\vec{r}') \psi(\vec{r}',t)
= 
\left( 
- \frac{\partial}{\partial t} 
+ \frac{1}{4m_N} \frac{\partial^2}{\partial t^2} 
\right)
\psi(\vec{r},t) 
\label{eq:Sch_2N:tdep}
\end{eqnarray}
even without the ground state saturation,
where
$\psi(\vec{r},t) \equiv G (\vec{r},t) / e^{-2m_N t}$
and 
$G (\vec{r},t)$ is a four-point correlation function,
$
%
G (\vec{r},t)
\equiv
\frac{1}{L^3}
\sum_{\vec{R}}
\langle 0 |
          (N(\vec{R}+\vec{r}) N (\vec{R}))(t)\
\overline{(N N)}(t=0)
| 0 \rangle
$ .
This ``time-dependent'' HAL QCD method is particularly useful
for nuclear systems, since the ground state saturation becomes
more and more difficult
at lighter quark masses and larger lattice volumes~\cite{HALQCD:2012aa}.

In practical lattice calculations,
it is difficult to handle the non-locality of the potential directly,
so we employ the derivative expansion of the potential, 
%
$
U(\vec{r},\vec{r}') =
\big[ V_C(r) + V_T(r) S_{12} + V_{LS}(r) \vec{L}\cdot \vec{S} + {\cal O}(\nabla^2) \big]
\delta(\vec{r}-\vec{r}') , 
$
%
%
where $V_C$, $V_T$ and $V_{LS}$ are the central, tensor and spin-orbit potentials, respectively,
with the tensor operator $S_{12}$.
In Ref.~\cite{Murano:2011nz}, the convergence of the derivative expansion is examined 
in parity-even channel,
and it is shown that the leading terms, $V_C$ and $V_T$, dominate the potential at low energies.

The HAL QCD method can be extended to systems above the inelastic threshold~\cite{Aoki:2011gt,Aoki:2012bb}
as well as multi-particle systems~\cite{Aoki:2013cra}.

\section{Lattice QCD setup and Numerical results}
\label{sec:results}

We employ
$N_f=2$ dynamical 
configurations
with mean field improved clover fermion 
and 
RG-improved
gauge action
generated by CP-PACS Collaboration~\cite{Ali Khan:2001tx}.
The measurements are performed at three different 
bare couplings $\beta=1.80, 1.95, 2.10$,
which corresponds to the lattice spacings
$a = 0.2150, 0.1555, 0.1076$ fm, respectively.
The physical lattice size is 
$L^3 \times T \simeq (2.5 {\rm fm})^3 \times 5 {\rm fm}$,
and 
the hadron masses
are 
$m_\pi \simeq 1.1$ GeV and $m_N \simeq 2.2$ GeV.
%
We use the wall quark source with Coulomb gauge fixing.
In order to enhance the statistics, we repeat the measurement 
using different source time slices and 
the forward and backward propagations are averaged.
Computational cost in the Wick and color/spinor contractions
are reduced by the unified contraction algorithm~\cite{Doi:2012xd}.
The simulation parameters are tabulated in Tab.~\ref{tab:lat_params}.
The cutoff dependence for $I=2$ $\pi\pi$ interaction
with the same configurations
has been investigated in Ref.~\cite{Yamazaki:2004qb}.

\begin{table}[b]
\begin{tabular}{ccccccccc} \hline\hline
$\beta$ & $L^3 \times T$    & $a$ [fm]    & $La$ [fm] & $\kappa_{ud}$ & $ m_\pi a$  & $m_N a$  & $N_{\rm conf}$ & $N_{\rm src}$ \\\hline
1.80    & $12^3 \times 24$  & 0.2150(22)  & 2.580(26) & 0.14090      & 1.1562(4)   & 2.262(2) &  640          & 24 \\
1.95    & $16^3 \times 32$  & 0.1555(17)  & 2.489(27) & 0.13750      & 0.8934(4)   & 1.695(1) &  598          & 32 \\
2.10    & $24^3 \times 48$  & 0.1076(13)  & 2.583(31) & 0.13570      & 0.6301(3)   & 1.182(1) &  798          & 48 \\\hline\hline
%
%
\end{tabular}
\caption{
Lattice simulation parameters.
$N_{\rm conf}$ is the number of configurations,
and $N_{\rm src}$ is the number of sources per each configuration in the measurement.
}
\label{tab:lat_params}
\end{table}

\begin{figure}[t]
\begin{minipage}{0.48\textwidth}
\begin{center}
\includegraphics[angle=0,width=1.0\textwidth]{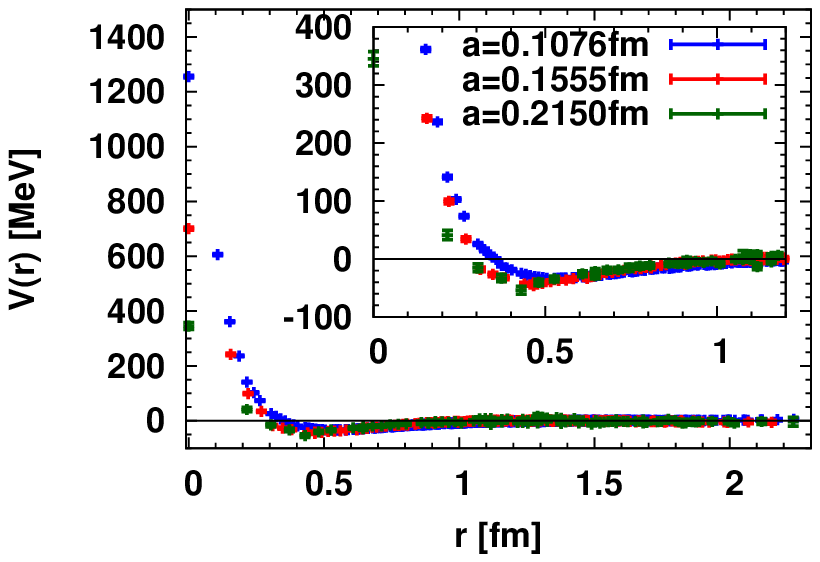}
\caption{
\label{fig:pot:1S0}
Nuclear central potential $V_C(r)$ in $^1S_0$ channel
obtained at three different lattice cutoffs.
}
\end{center}
\end{minipage}
\hfill
\begin{minipage}{0.48\textwidth}
\begin{center}
\includegraphics[angle=0,width=1.0\textwidth]{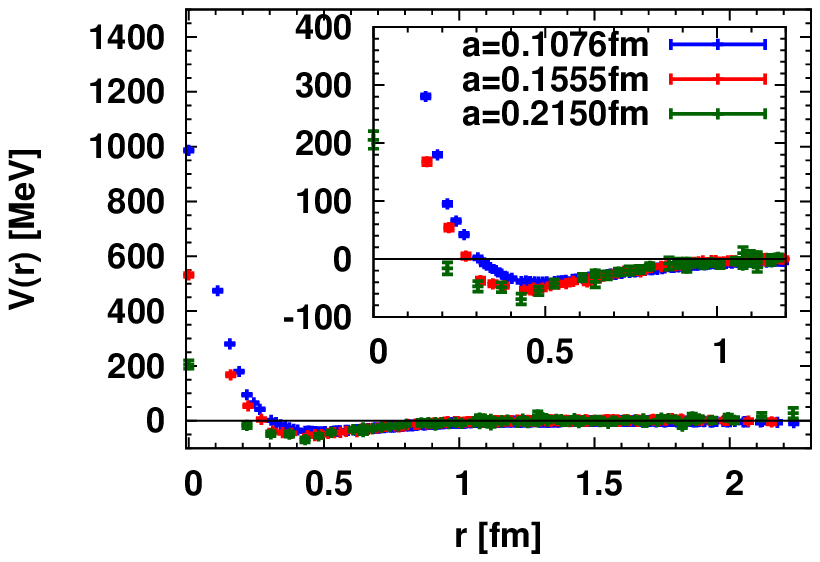}
\caption{
\label{fig:pot:3S1-3D1:cen}
Nuclear central potential $V_C(r)$ in $^3S_1$--$^3D_1$ channel.
}
\end{center}
\end{minipage}
\end{figure}

\begin{figure}[t]
\begin{minipage}{0.48\textwidth}
\begin{center}
\includegraphics[angle=0,width=1.0\textwidth]{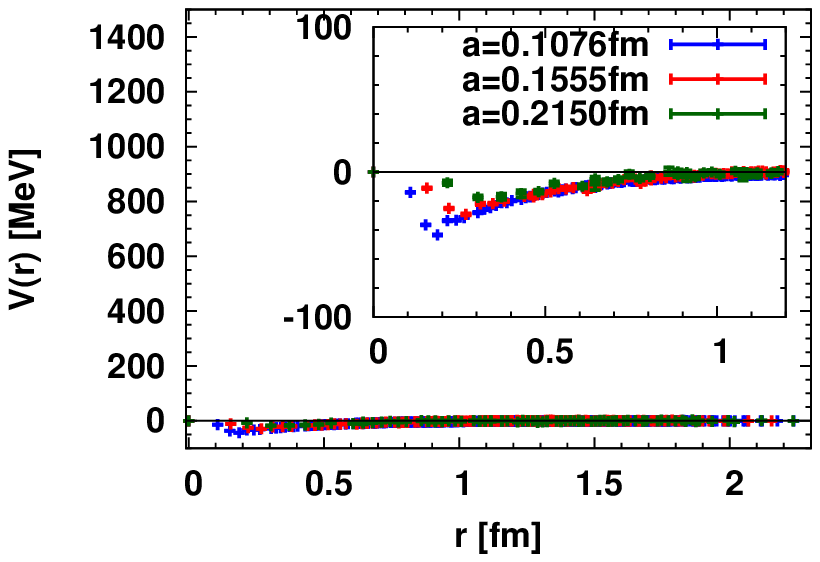}
\caption{
\label{fig:pot:3S1-3D1:ten}
Nuclear tensor potential $V_T(r)$ in $^3S_1$--$^3D_1$ channel.
}
\end{center}
\end{minipage}
\hfill
\begin{minipage}{0.48\textwidth}
\begin{center}
\includegraphics[angle=0,width=1.0\textwidth]{%
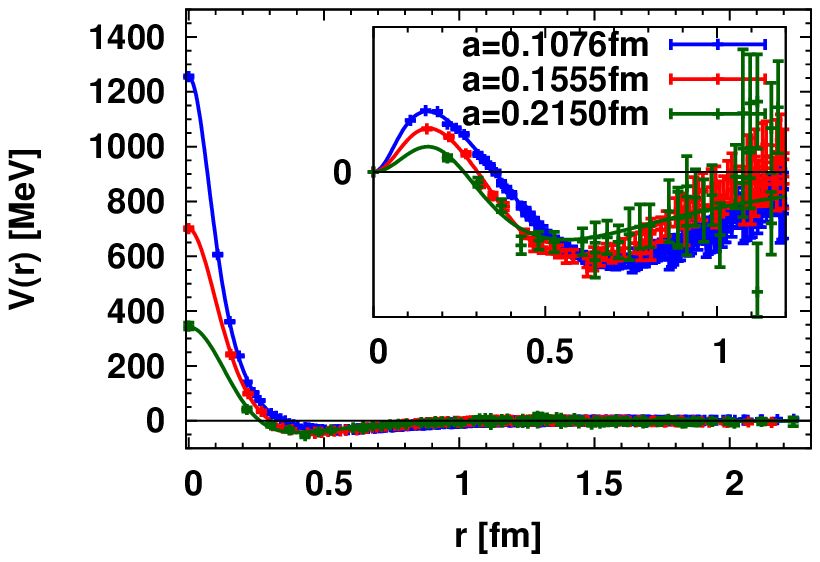}
\caption{
\label{fig:pot_r2:1S0}
The fit of 
$V_C(r)$ in $^1S_0$ channel.
The inner figure shows $r^2 V_C(r)$ to include phase space factor.
}
\end{center}
\end{minipage}
\end{figure}

In Fig.~\ref{fig:pot:1S0}, we plot the nuclear central potential in $^1S_0$ channel
for each lattice cutoff.
In the evaluation of Eq.~(\ref{eq:Sch_2N:tdep}),
we omit the second derivative term in $t$,
which corresponds to the relativistic correction
and is expected to have marginal effect at heavy quark mass~\cite{HALQCD:2012aa}.
Shown in Figs.~\ref{fig:pot:3S1-3D1:cen} and ~\ref{fig:pot:3S1-3D1:ten} 
are the central and tensor potentials in $^3S_1$--$^3D_1$ channel, respectively.

Comparing the results at three different cutoffs,
we observe non-negligible cutoff dependence 
at the short distance part of the potentials,
while the cutoff dependence at long distance region is suppressed.
This is a natural consequence that the discretization effect has
intrinsically short-range nature.
We note that
a similar observation is obtained in 
the potential of $I=2$ $\pi\pi$ system,
where the violation of rotation symmetry due to finite lattice spacing 
is found at short distance~\cite{Kurth:2013tua}.

It is also interesting that repulsive cores in central forces 
are enhanced on a finer lattice.
This tendency is consistent with the 
analyses by the operator product expansion (OPE),
where the repulsive cores are predicted to diverge in the limit of  $r\rightarrow 0$~\cite{Aoki:2010kx}.
We, however, note that 
even the finest lattice ($a=0.1076$ fm) may not be sufficiently fine to 
quantitatively examine these perturbative behaviors.
In the tensor force, 
we observe kink structures at $r\sim$ 0.2--0.3 fm (depending on $a$),
and 
the results from different cutoffs
agree if the distance $r$ is longer than the kink position,
while the deviation is observed at shorter distance than the kink.
In fact, we have been observing similar kink structures in tensor forces
in various baryon-baryon interactions with a variety of lattice setup~\cite{Aoki:2012tk}.
The systematic study in this report indicates that
these kink structures are associated with the discretization artifact.

While the lattice cutoff dependence in potentials looks sizable at short distance,
it is important to realize that such effect is expected to be suppressed 
in physical observables such as phase shifts and scattering length,
because of the phase space factor of $\propto r^2$.
In order to make a quantitative study, we calculate the 
phase shifts and scattering length in $^1S_0$ channel.
In Fig.~\ref{fig:pot_r2:1S0}, we show the potential together with 
their fitted values where we employ the fitting function used in~\cite{Inoue:2011ai}.
Shown in the inner figure is the potential multiplied by $r^2$ to account for the phase space factor.
One can observe that the discretization artifact is effectively suppressed as discussed above.

\begin{figure}[t]
\begin{minipage}{0.48\textwidth}
\begin{center}
\includegraphics[angle=0,width=1.0\textwidth]{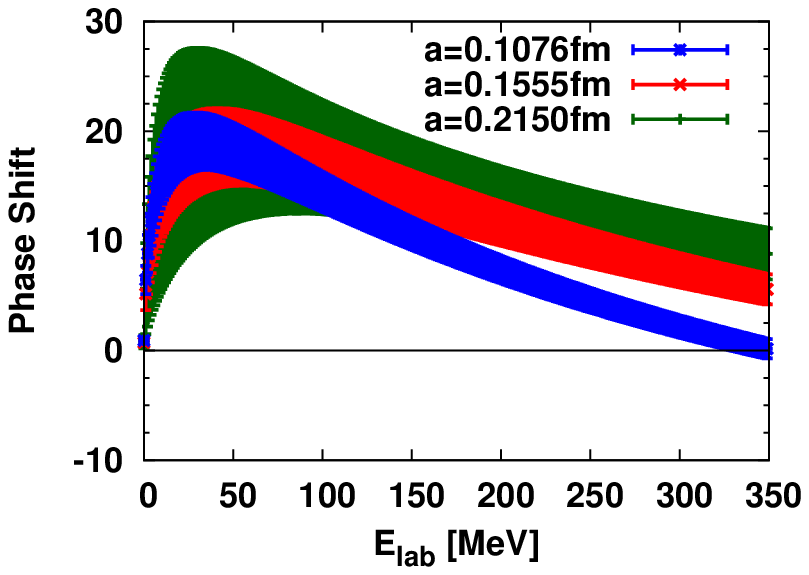}
\caption{
\label{fig:phase:1S0}
Phase shifts in $^1S_0$ channel in terms of the laboratory energy.
}
\end{center}
\end{minipage}
\hfill
\begin{minipage}{0.48\textwidth}
\begin{center}
\includegraphics[angle=0,width=1.0\textwidth]{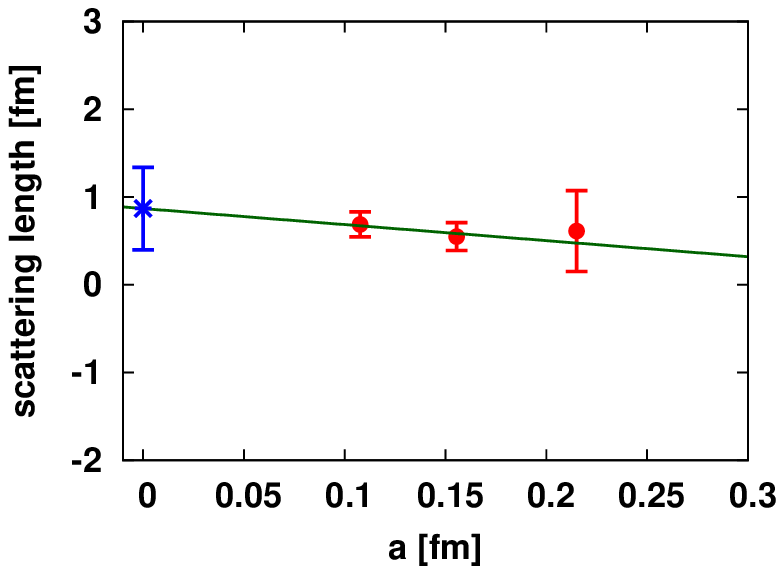}
\caption{
\label{fig:scatt_len:1S0}
The scattering length in $^1S_0$ channel 
against the lattice spacing $a$.
The blue point corresponds to the result in the continuum limit
obtained by the linear extrapolation against $a$.
}
\end{center}
\end{minipage}
\end{figure}

By solving the Schr\"odinger equation 
using the fitted potential in infinite volume,
we obtain the phase shifts.
In Fig.~\ref{fig:phase:1S0}, we plot the phase shifts 
in terms of the laboratory energy
where the bands correspond to the statistical fluctuations.
The qualitative behaviors are found to be similar to the experimental phase shifts.
We observe that the results from different cutoffs agree at low energies
within the statistical errors.
On the other hand, there exists a deviation at high energies,
where phase shifts become smaller on a finer lattice.
These behaviors can be understood 
by recalling that
the cutoff dependence appears only at short range part of the potential,
and the repulsive core is enhanced on a finer lattice.
In Fig.~\ref{fig:scatt_len:1S0}, we show preliminary results for the 
scattering length
$a(^1S_0) = \lim_{k\rightarrow 0} \tan\delta(k)/k$
against the lattice spacing $a$,
where the error is statistical only.
Since the scattering length represents the low-energy phenomena,
the cutoff dependence is found to be negligible compared to the statistical errors.
Detailed studies for the systematic uncertainties 
for phase shifts and scattering length
are in progress.

We thank authors and maintainers of CPS++\cite{CPS},
whose modified version is used in this study. 
We also thank  
CP-PACS Collaboration
and ILDG/JLDG~\cite{conf:ildg/jldg} for providing gauge configurations.
The numerical simulations have been performed
on SR16000 and Blue Gene/L at KEK,
T2K at University of Tsukuba, SR16000 at YITP in Kyoto University
and FX10 at University of Tokyo.
This research is supported in part by MEXT Grant-in-Aid 
for Scientific Research (No. 24740146),
for Scientific Research on Innovative Areas (No. 2004: 20105001, 20105003)
and SPIRE (Strategic Program for Innovative REsearch).


\end{document}